\documentclass[fleqn,twoside]{article}
\usepackage{espcrc2}
\usepackage{epsf}
\input{psfig}
\input{epsf}
\usepackage{epsfig}
\usepackage{psfig}
\newcommand{\be}{\begin{equation}}
\newcommand{\ee}{\end{equation}}

\title{ Running coupling constant and propagators
        in $SU(2)$ Landau gauge\thanks{Talk presented by A. Cucchieri}}
\author{Jacques C.\ R.\ Bloch\address[TUEBINGEN]{Institut f\"ur Theoretische 
        Physik, Universit\"at T\"ubingen, D-72076 T\"ubingen, Germany},
        Attilio Cucchieri\address[IFSC]{IFSC S\~ao Paulo University,
        C.P. 369 CEP 13560-970, S\~ao Carlos (SP), Brazil}\thanks{
        Research supported by
        FAPESP, Brazil (Project No.\ 00/05047-5)},
        Kurt Langfeld\addressmark[TUEBINGEN]
        and Tereza Mendes\addressmark[IFSC]$^{\dag}$}
\begin{document}

\begin{abstract}
We present a numerical study of the running coupling constant
and of the gluon and ghost propagators in
minimal Landau gauge. Simulations are done in pure
$SU(2)$ lattice gauge theory for several values of $\beta$
and lattice sizes. We use two different lattice setups.
\end{abstract}
\maketitle

\section{INTRODUCTION}

We consider, on the lattice, a running coupling
constant $g^{2}(p)$ defined by \cite{vS,B}
\begin{equation}
g^{2}(p) \equiv g^{2}_{0}\,\left[\,p^2\,D(p)\,\right]
\,\left[\,p^2\,G(p)\,\right]^2
\label{eq:alphadef}
\end{equation}
where $D(p)$ and $G(p)$ are, respectively, the~gluon and ghost
propagators evaluated~in Landau gauge.
Clearly $g^{2}(p)$ is a gauge-dependent quan\-tity; however,
notice that $g^{2}(p)$ is re\-nor\-ma\-li\-za\-tion-group invariant
in Landau gauge since, in this case,
$Z_g Z_3^{1/2} \widetilde{Z_3}
=\widetilde{Z_1} = 1 $.
This running coupling strength enters the quark
Dyson-Schwinger equation directly and can be
interpreted as an effective interaction strength
between quarks \cite{alpha}.

Studies of the coupled set
of Dyson-Schwinger equations for the gluon
and ghost propagators have shown that: (i)
the gluon propagator behaves as
$D(p) \sim p^{-2+4 \kappa}$ in the infrared
limit [and thus $D(0) = 0$ if $\kappa > 0.5$],
(ii) the ghost propagator behaves as
$G(p) \sim p^{-2-2 \kappa}$ at small momenta
and (iii) the running coupling strength
$\alpha_s(p) = g^2(p) / 4 \pi$
defined in eq.\ (\ref{eq:alphadef})
has a finite value $\alpha_c$ at zero momentum
(infrared fixed point).
Using different approximations, in order to solve the
Dyson-Schwinger equations, the following values have
been obtained: $\kappa \approx 0.92$ and $\alpha_c 
\approx 9.5$ \cite{vS}, $\kappa \approx 0.77$
and $\alpha_c \approx 11.5$ \cite{B}, $\kappa \approx 
0.60$ and $\alpha_c \approx 8.9/N_c$ \cite{newk}.
[Here, the first two results refer to $SU(3)$.] We stress
that the large value for $\alpha_c$ obtained in \cite{vS,B}
is related to the angular approximation used in the
integration kernels. Let us notice that, using
stochastic quantization \cite{DanStoc}, Zwanziger 
also obtained that the transverse gluon propagator
in the infrared limit behaves as $D(p)
\sim p^{-2+4 \kappa}$ with $\kappa \approx 0.52$. 

From the lattice point of view we
know that lattice gauge-fixed Landau configurations
belong to the region $\,\Omega\,$
delimited by the first Gribov horizon, and that
$\Omega$ is not free of Gribov copies.
One can also prove \cite{DanVanish} that the restriction of
the path integral to the region $\Omega$ implies a
suppression of the (unrenormalized) transverse gluon
propagator $D(p)$ in the infrared limit. At the same time,
the Euclidean probability gets concentrated
near the Gribov horizon and this implies enhancement
of $G(p)$ at small momenta \cite{DanFMR}.

\section{RESULTS} 

Simulations have been done in S\~ao Carlos
for $\beta = 2.2, 2.3, \ldots, 2.8$
and $V = 14^4$, $20^4$, $26^4$,
and in T\"ubingen for $\beta =
2.1, 2.15, \ldots, 2.5$
and $V = 12^3 \times 24$, $16^3 \times 32$.
The simulations carried out in T\"ubingen are based
on a direct evaluation of the form 
factors $F(p) = D(p)\, p^2$ and $\,G(p)\, p^2\,$
appearing in eq.\ (\ref{eq:alphadef}).
Also, for the evaluation of $F(p)$,
the gluon field has been defined in terms of
the adjoint links  \cite{Kurt} instead of the
usual link variables. The gluon field obtained in this
way is invariant under non-trivial $Z_2$ transformations.

Gribov-copy effects for the two propagators, if present,
are smaller than the numerical accuracy \cite{Kurt,A}.
Preliminary results have been presented in \cite{Kurt,ATD}.

\vskip 3mm

In order to compare lattice data obtained for the
two propagators at different $\beta$ values we
used a standard scaling analysis \cite{scaling}
based on maximum overlap without considering
any phenomenological fit functions.
(Details will be presented in \cite{BAKT}.) Also,
for the data produced in S\~ao Carlos, we have discarded
data points at small momenta that are affected by
finite-size effects. (These finite-size effects are
less pronounced when one evaluates the form factor
directly.)

\begin{figure}[t]
\begin{center}
\epsfxsize=0.36\textwidth
\leavevmode\epsffile{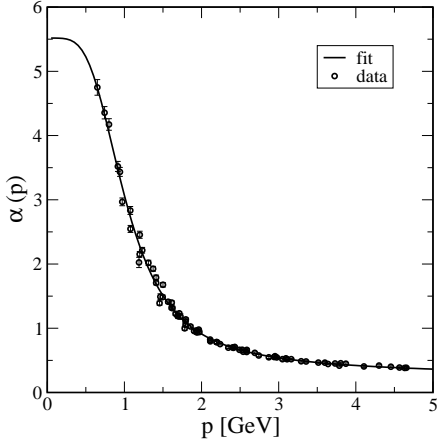}
\vspace*{-0.8cm}
\caption{Fit for the running coupling using eq.\
(\ref{eq:alphaT}) with
$c_0 = 1.4(2)$, $a_0 = 5.5(3)$, $\delta = 1.77(9)$,
$\Lambda = 0.83(4)$ and $\lambda$ set to $2.2$.
\vspace*{-0.7cm}}
\label{fig:alphaT}
\end{center}
\end{figure}

We have considered two different sets of fitting
functions, namely
\begin{eqnarray}
\alpha (p) \!\!\!&=&\!\!\!  \frac{1}{c_0+t^\delta } \,
\Bigl[ c_0 a_0 \; + \; \alpha _2 (t+\lambda ) \; t^\delta \Bigr]
\label{eq:alphaT} \\
D(p)\, p^2 \!\!\!&=&\!\!\! A \, \frac{t}{c_1 \; + \; c_2 \, t^\frac{1}{2} + t}
\,\alpha ^{13/22} (p)
\label{eq:DT} \\
G(p) \,p^2 \!\!\!&=&\!\!\! B \Bigl( \frac{c_1 \; + \; c_2 \, t^\frac{1}{2} +t}{t}
\Bigr)^\frac{1}{2} \,
\alpha ^{9/44} (p)
\label{eq:GT}
\end{eqnarray}
where $t=p^2/\Lambda^2$ and $\alpha_2(p)$ is the 2-loop running
coupling constant \cite{2loop}, and
\begin{eqnarray}
\alpha(p) \!\!\!&=&\!\!\! C \, p^4 / \left[ (p^4 + m) \, s (a) \right]
\label{eq:alphaSC} \\
D(p) \!\!\!&=&\!\!\! A\, p^2 / \left[ (p^4 + m) \, s^{\gamma_D}(a_D) \right] 
   \label{eq:DSC} \\
G(p) \!\!\!&=&\!\!\! B\, / \, \left[p^2 \,  s^{\gamma_G}(a_G) \right] 
   \label{eq:GSC} 
\end{eqnarray}
where $s(a) = (11 / 24 \pi^2) \log{[1 + (p^2/\Lambda^2)^{a}]}$,
$\gamma_D = 13/22$ and $\gamma_G = 9/44$.
Note that, in the first case, the fitting functions correspond to
$\kappa = 0.5$, while in the second case one has
$\kappa_G = a_G \gamma_G$ and $\kappa_D = 1 - a_D \gamma_D / 2$.
Also, both sets of fitting functions satisfy the
leading ultraviolet behavior of the two propagators.

\begin{figure}[t]
\begin{center}
\epsfxsize=0.4\textwidth
\leavevmode\epsffile{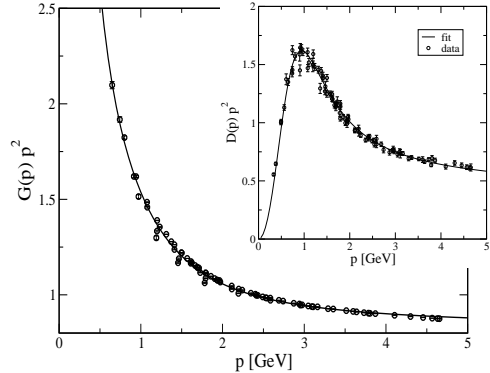}
\vspace*{-0.8cm}
\caption{Fit for the ghost and gluon propagator form factors
using eqs.\ (\ref{eq:GT}) and (\ref{eq:DT}) respectively, with
$c_1 = 0.98(4)$, $c_2=-0.59(6)$, $A =0.98(2)$, $B = 1.124(9)$
and $\alpha(p)$ as obtained from the fit reported in 
Fig.\ \ref{fig:alphaT}.
\vspace*{-0.9cm}}
\label{fig:DT}
\end{center}
\end{figure}

\begin{figure}[t]
\vspace*{-2.9cm}
\begin{center}
\epsfxsize=0.43\textwidth
\leavevmode\epsffile{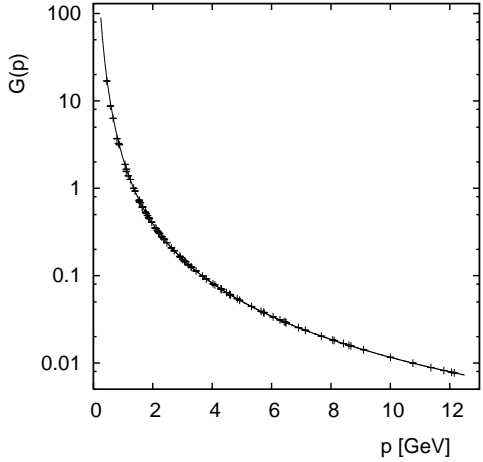}
\vspace*{-0.8cm}
\caption{Fit for the ghost propagator
using eq.\ (\ref{eq:GSC}) with
$B = 0.924(4)$, $a_G = 1.73(3)$
and $\Lambda = 1.322(8)$;
this gives $\kappa_G = a_G \gamma_G = 0.354(6)$.
If $\gamma_G$ is also a fitting parameter we get
$\gamma_G = 0.202(5)$ to be compared with
$9/44 \approx 0.2045$.
\vspace*{-0.8cm}}
\label{fig:GSC}
\end{center}
\end{figure}

\begin{figure}[t]
\vspace*{-4.4cm}
\begin{center}
\epsfxsize=0.43\textwidth
\leavevmode\epsffile{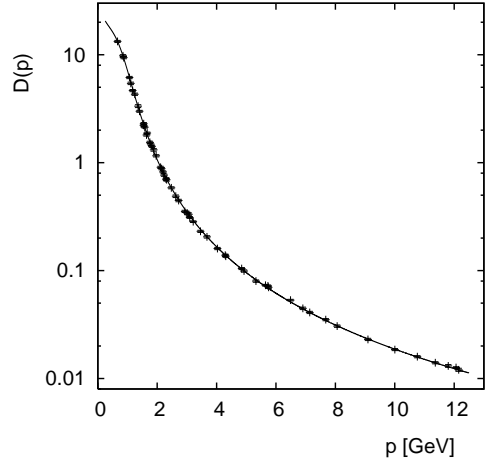}
\vspace*{-0.8cm}
\caption{Fit for the gluon propagator
using eq.\ (\ref{eq:DSC}) with
$A = 1.02(9)$, $a_D = 1.9(3)$ and $m = 0.8(3)$;
this gives
$\kappa_D = 1 - a_D \gamma_D / 2 = 0.44(9)$.
Here $\Lambda$ has been set to $1.322$ (see
Fig.\ \ref{fig:GSC}).
If $\gamma_D$ is also a fitting parameter we get
$\gamma_D = 0.579(7)$ to be compared with
$13/22 \approx 0.591$.
\vspace*{-1.0cm}}
\label{fig:DSC}
\end{center}
\end{figure}

\begin{figure}[htb]
\vspace*{-4.4cm}
\begin{center}
\epsfxsize=0.43\textwidth
\leavevmode\epsffile{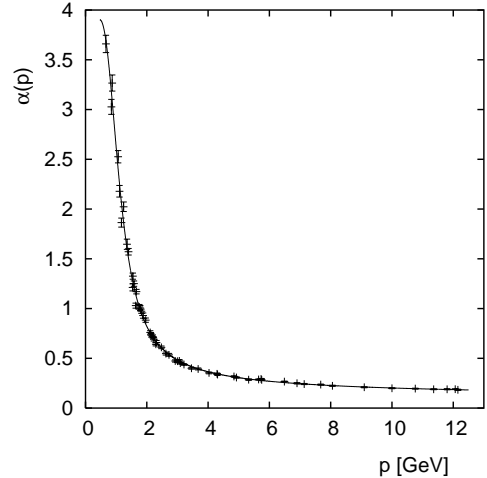}
\vspace*{-0.8cm}
\caption{Fit for the running coupling $\alpha(p)$
using eq.\ (\ref{eq:alphaSC}) with
$C = 0.072(8)$,
$a = 1.9(3)$, $\Lambda = 1.31(1)$
and $m = 1.0(6)$.
\vspace*{-1.0cm}}
\label{fig:alphaSC}
\end{center}
\end{figure}

\vskip 3mm

\setcounter{footnote}{0}
Results of the fits are reported\footnote{Notice
the logarithmic scale in the $y$ axis in 
Figs.\ \protect\ref{fig:GSC}, \protect\ref{fig:DSC}.}
in Figs.\ \ref{fig:alphaT}--\ref{fig:alphaSC}. 
From our data there is evidence for the
suppression of the transverse gluon propagator $D(p)$
in the infrared limit and for the enhancement of
the ghost propagator $G(p)$ in the same limit.
Also, the running coupling strength $\alpha_s(p)$
defined in eq.\ (\ref{eq:alphadef}) probably has a finite
value at zero momentum. However, in order to
probe the infrared region and give a final value for
$\kappa$ and $\alpha_c$ one needs to simulate at larger
lattice volumes.

\end{document}